\documentclass[manuscript]{aastex}
\usepackage{natbib}
\usepackage{graphicx}
\usepackage{dcolumn}
\usepackage{bm}
\usepackage{multirow}
\usepackage{amsmath}
\usepackage{amsfonts}

\shorttitle{AK Sco }
\shortauthors{G\'omez de Castro et al}

\begin{document}

\title{XMM-Newton monitoring of the close pre-main-sequence binary AK Sco. Evidence of tide driven filling of the inner gap in the circumbinary disk}
\author{Ana In\'es G\'omez de Castro}
\affil{S.D. Astronom\'ia y Geodesia and Instituto de Matem\'{a}tica Interdisciplinar, Fac. de CC Matem\'{a}ticas, Universidad Complutense, 28040 Madrid, Spain}

\author{Javier L\'opez-Santiago}
\affil{Departamento de Astrof\'isica, Fac de CC F\'isicas, Universidad Complutense,
28040 Madrid, Spain}
\author{Antonio Talavera}
\affil{European Space Astronomy Center, Madrid, Spain}
\author{A.Yu. Sytov}
\affil{Institute of Astronomy of the Russian Academy of Sciences, Moscow, Russia}
\author{D. Bisikalo}
\affil{Institute of Astronomy of the Russian Academy of Sciences, Moscow, Russia}

\begin{abstract}
AK~Sco stands out among pre-main sequence binaries because of its prominent ultraviolet excess, the high eccentricity of its orbit and the strong tides driven by it. AK Sco consists of two F5 type stars that get as close as 11R$_*$ at periastron passage. The presence of a dense
($n_e \sim 10^{11}$~cm$^{-3}$) extended envelope has been unveiled recently. In this article, we report the results from a XMM-Newton based monitoring  of the system. We show that at periastron, X-ray and UV fluxes are enhanced
by a factor of $\sim 3$ with respect to the apastron values. The X-ray radiation is produced in an optically thin plasma with T$\sim 6.4\times 10^{6}$ K and  it is found that the N$_H$ column density rises from 0.35$\times 10^{21}$~cm$^{-2}$ at periastron to 1.11$\times 10^{21}$~cm$^{-2}$ at apastron, in good agreement with previous polarimetric observations.  The UV emission detected in the Optical Monitor band seems to be caused by the reprocessing of the high energy magnetospheric radiation on the circumstellar material. 
Further evidence of the strong magnetospheric disturbances is provided by the detection of line broadening of 278.7~km~s$^{-1}$ in the N~V line with HST/STIS. 

Numerical simulations of the mass flow from the circumbinary disk to the components have been carried out. They provide a consistent scenario with which to interpret AK~Sco observations. We show that the eccentric orbit acts like a gravitational piston. At apastron, matter is dragged  efficiently from the inner disk border, filling the inner gap and producing accretion streams that end as ring-like structures around each component of the system. At periastron, the ring-like structures come into contact, leading to angular momentum loss, and thus producing an accretion outburst.

\end{abstract}

\section{Introduction}

Late type pre-main sequence stars, T Tauri Stars (TTSs), are characterised by their strong radiation excess at high energies, from UV to X-ray wavelengths. There has been
much discussion regarding the source of this excess. In the 80's and early 90's it was thought that this excess was associated with an enhancement of the magnetic activity caused by the deep convective layers of the TTSs
prior to stellar stabilisation on the main sequence (Giampapa et al. 1981, Calvet et al., 1985). Later on, the realisation that accretion itself could be the source of energy, drove the development of several models where both UV and X-ray excesses were produced by accretion shocks; material free-falling along the stellar field lines onto a shock front at the stellar surface (Lamzin 1998, G\'omez de Castro \& Lamzin 1999, Gullbring et al 2000, Guenther et al 2007). The shock front temperature may reach a few $10^{6}$ K, hence producing soft X-ray emission and UV excess radiation. In 2001, it was realised that outflows from these systems are hot enough to contribute to the UV excess (G\'omez de Castro \& Verdugo, 2001); the magnetically mediated interaction between the keplerian accretion disk and the star basically transforms angular momentum from the disk storage into mechanical energy (outflows) and heating. The phenomenon is non-stationary and it is controlled by the rotation period and the diffusion time scale. There is general consensus on the co-existence of two components to this outflow: a warm high density component associated with a centrifugally driven wind from the disk, and a hot low density component associated with plasmoid ejection from the magnetic interface between the stellar field and the disk (von Rekowski \& Brandenburg 2004, G\'omez de Castro \& von Rekowski, 2011). Thus, jets are expected to have a non-stationary, magnetised, hot component. Indeed, X-ray emission has been detected from some protostellar jets (Favata~et~al.~2002; Bonito et al. 2007; Guedel~et~al.~2008).

Current understanding of the physics of star formation indicates that magnetic activity, accretion and
outflow, are contributing to the high energy excess of the pre-main sequence (PMS) systems. The stellar magnetosphere acts as the interface between the various dynamical components (disk, star) and channels/dials the flows driven by them (accretion and winds/jets).

The magnetospheres of PMS stars are very powerful and extend 2-8 stellar radii to the limits of the inner disk boundary (Johns-Krull 2007, G\'omez de Castro \& Marcos-Arenal 2012, hereafter GdCMA2012). They  are powered
by the stellar magnetic field, the accreting infalling material, and shearing at the boundary between the magnetosphere
and the inner disk (see Hartmann 2009 for a recent review and G\'omez de Castro 2009a for the UV radiative output of this
engine). Magnetospheric heating processes during PMS evolution are poorly studied, as well as the relative relevance of the contributions mentioned above. Close PMS binaries provide an ideal laboratory to study magnetospheric physics since the coupling between the stellar magnetospheres and the disk is broken and gravity controls the mass transfer from the inner border of the disk to the stars. Numerical simulations predict the dynamical clearing of the inner region of the circumbinary disk (Artymowicz \& Lubow 1994; Bate \& Bonnel 1997, Kaigodorov et al 
2010, Fateeva et al 2011, de Val-Borro et al 2011). Indeed, recent infrared studies have shown that most of the
(few) known PMS spectroscopic binaries (PMS-SBs) (Mathieu et al 1997 and references therein) do not display significant infrared excess in the 1-5$\mu m$ range suggesting cleared regions within the circumbinary disk (Jensen \& Mathieu 1997). The size of the gaps is consistent with the theoretical predictions for dynamically clearing. However,  PMS-SBs display UV excesses as high as their single counterparts do (enhanced emission of the so-called chromospheric/transition region tracers). This suggests that the disk-star locking is not the dominant source of magnetospheric heating, rather it seems to be controlled by the stellar activity and the accretion flow.

AK Sco stands out among the PMS-SBs because of the evidence of transient material filling the gap and its prominent ultraviolet excess. The high eccentricity of AK~Sco's orbit triggers strong tides in the inner border of the circumbinary disk as well as in both components;  AK Sco is made of two F5 type stars that get as close as 11R$_*$ at periastron passage (see Table~1 for a summary of AK~Sco properties). Moreover recent UV observations have unveiled the presence of dense and extended magnetospheres around
each component (G\'omez de Castro, 2009b). Each stellar magnetosphere extends to $\sim 1.7 $R$_*$ and is highly turbulent; the region is disturbed by a velocity field ($\sigma  \simeq100$~km~s$^{-1}$) that exceeds significantly the sound  speed in the UV radiating plasma. The radiative loses of the magnetospheres (0.04L$_{\odot}$ per star) cannot be accounted solely by the dissipation of energy from the tidal wave hence, the contribution of the accretion flow needs to be taken into account. However, no signs of enhanced accretion at periastron passage have been reported from the optical H$\alpha$/H$\beta$ profiles (Andersen et al. 1989, hereafter A1989; Alencar et al. 2003, hereafter A2003). This is not the generic case for PMS-SBs; for instance, KH 15D, shows phase dependent variations of its H$\alpha$ profiles, which have been interpreted as a sign of an accretion stream member (Hamilton et al., 2012).

In this article, we present evidence of the enhancement of energy release at periastron based on the X-ray/UV monitoring of AK~Sco with the XMM-Newton space telescope. The presence of extended structures around each component of the binary system is confirmed by recent observations of AK~Sco obtained from the Hubble Space Telescope (HST) Archive. Also evidence of an enhanced dragging of material from the inner disk border into the gap at apastron, is reported. Numerical simulations of the stars-disk interaction have been carried out to analyse the observations. It is shown that the high eccentricity of AK~Sco orbit acts like a gravitational piston that efficiently drags matter from the disk border at apastron, feeding an enhanced mass infall at periastron. In Sect.~2, we describe the XMM-Newton observations and data reduction as well as the Archival data obtained from the International Ultraviolet Explorer (IUE) and the Hubble Space Telescope (HST) . In Sect.~3, we present the analysis of these data and the obtained results. 
The binary-disk model and the numerical set-up developed to study the evolution of the system is described in Sect.~4.. In Sect.~5 we propose the general framework to interpret the observations  Finally, a brief summary is provided in Sect.~6. 

\section{Observations and data reduction}

\subsection{XMM-Newton Observations and data reduction}

The XMM-Newton monitoring of AK Sco was carried out from March 15th, 2011 to March 22nd, 2011. The observations were performed at phases\footnote{The phases are calculated using the ephemeris by A2003 with phase 0 corresponding to periastron passage.} 0.99, 0.15 and 0.48, corresponding to observation identifications, ID. 0651870201, ID. 0651870301 and ID. 0651870401, respectively. The observing log is presented in Table 2.

The data obtained in the three XMM-Newton observations have been reduced using the standard Science Analysis System (SAS) developed by the XMM-Newton project.
SAS allows the user to apply all instrumental corrections and calibrations corresponding to the instruments on board XMM-Newton. SAS version 11 has been used
for the reduction of both EPIC and OM data.

\subsubsection{EPIC data}

The European Photon Imaging Cameras (EPIC) were used in each observation in full-frame mode 
with the thick filter for an exposure time of approximately 25 ks. The standard procedure was followed for EPIC data reduction and calibration. Firstly, calibrated file events were created using SAS tasks
\textit{epproc} and \textit{emproc} for each observation independently.

Light curves obtained from events detected in the whole field of view with energy $E > 10$keV were explored to identify high background periods which did not appear in any of the three observations.
The event lists were then cleaned for bad events and noise. The resultant event files were used for the extraction of source light curves and spectra of AK Sco.

\subsubsection{Optical and UV Monitor Observations and Data Reduction}

The Optical and UV Monitor (OM) was operated in User Defined Mode, with the
UVM2 filter (effective wavelength $\sim$ 2310 \AA ). Four exposures of 4400s were obtained in each
observation. Two detector windows were defined in each exposure, an image mode
window of 5$\times$5 arcmin$^2$ and a fast mode one of 12$\times$12 arcsec$^2$ centred on AK~Sco. 
The fast mode window allowed us to obtain data with a time resolution of 1 s.

Aperture photometry has been performed on the OM images, obtained with the
UVM2 filter. The results obtained with SAS, corrected count rates and the
corresponding AB magnitudes are given in Table~3. 


Detailed light curves have been derived from the fast mode OM event data (see Section 3.1).
In order to increase the signal to noise ratio, a time binning of 100 s has
been used. Special care has
been taken to check that the target was well centred in the small fast mode
window in all exposures and to ensure that the background was properly
measured.

\subsection{UV Archival data of AK~Sco}

The UV spectra of AK~Sco obtained with the International Ultraviolet Explorer (IUE) have been retrieved from the IUE Newly Extracted Spectra (INES) archive for this study. The log of observations is in Table~4 (see G\'omez de Castro \& Franqueira 1997 for further details). IUE observations were obtained with the Long Wavelength Prime camera (LWP), Long Wavelength Redundant Camera (LWR) and the Short Wavelength Prime camera (SWP) in low dispersion mode. The long wavelength cameras obtained the spectrum in the 2000 \AA - 3200 \AA\ spectral range and the SWP camera in the 1200 \AA - 1950 \AA\ range. IUE was equipped with a Fine Error Sensor (FES) that was also used to provide estimates of the visual magnitude with RMS errors of 0.08 mag (Holm and Rice 1981).  The FES measurements have been converted into magnitudes using Stickland's (1980) and Perez's (1991) calibration. They have been corrected for colour effects,
the sensitivity degradation of the cameras (Fireman and Imhoff 1989) and the change of the FES reference point after January and July 1990 at the two ground control stations, NASA and European, respectively. FES magnitudes are also provided in Table~4. AK~Sco was observed with the IUE in low dispersion (6 \AA\ resolution) and in high dispersion (0.2 \AA\ resolution) modes. The binary was too weak to be observed in the high dispersion mode below $\sim 2700$ \AA . As shown in Table~4, exposure times were of 6-9 hours. Only the Mg~II lines are observed with a good enough signal-to-noise ratio (SNR) (see Section 3.5).

\subsection{HST/STIS Archival Data of AK~Sco}

Recently, high SNR spectra of AK~Sco have been obtained making use of the Space Telescope Imaging Spectrograph  (STIS) on board the Hubble Space Telescope (HST). The observations were obtained in echelle mode with gratings E230M
(R $\sim$ 60,000) and E140M (R$\sim 91,000$) on 2010 August 21 at phase 0.84. The spectra cover the range 1150~\AA - 3150~\AA\ in three observations obtained sequentially (see Table 4 for further details).

\section{Data Analysis}

\subsection{UV photometry and light curves}
\label{sec:UVvariability}

The results of the photometry with OM are displayed in Figure~1 where we plot the AB magnitudes in the UVM2 filter (see Talavera, 2011, for OM calibration). We plot here as well the integrated fluxes obtained from IUE spectra in the OM UVM2 band, also as AB magnitude. All these data are overimposed in the radial velocity curve from A2003. We see clearly how the fluxes are higher at periastron (phase 0.0), both in the OM and IUE data (note that IUE data were not obtained in the same binary system cycle than OM data). The OM band is sensitive to UV emission in the 2100\AA - 2800\AA\ spectral range.
AK~Sco spectrum in this band  is dominated by the Balmer continuum and the Fe~II resonance multiplets (i.e. by plasma with an electron temperature of about 10$^4$~K; the so-called veiling spectrum); there are not strong spectral lines and the photospheric contribution is negligible. In Figure~2, the effective area of the XMM-OM in the UVM2 band has been multiplied by the UV spectrum of AK~Sco (from IUE data, see Sect.~3.3) at two phases, showing the higher flux of the star at periastron, depicted also in Figure~1. Note that this veiling spectrum is a good indicator of the accretion rate in the T Tauri stars (Ingleby et al. 2011).

The light curves from the three XMM-Newton observations with OM in fast mode are shown in Figure~3,
where the abscissa is referred to the start of each observation. In addition to the variability with the phase shown in 
Figure~1 we see in the fast mode some short time scale variation along the duration of each observation.
This short term variability is analysed in G\'omez de Castro et al (2012).

\subsection{Modeling the X-ray spectrum of AK~Sco}

Spectra of the X-ray source were extracted from circles with radius 6 arcmin
(this region contains 90\% of the source photons, on-axis). Background spectra
were extracted from a nearby region of the same chip, free of any other X-ray
source. Response matrices and ancillary files were created for each observation.
The analysis of the spectra was done with XSPEC\footnote{http://heasarc.nasa.gov/xanadu/xspec/}
(Arnaud 1996, 2004). Independently  for each observation, the background-corrected source
spectrum was fit against a hot plasma model. We chose
the model by Smith et al. 2001a, that uses the atomic data contained in the Astrophysical
Plasma Emission Database (APED, Smith et al. 2001b) and it is available in XSPEC.
A multiplicative interstellar absorption model --\,in particular the
one described in Morrison \& McCammon (1983),-- was added to account for the possible
effects of interstellar or circumstellar material of the source X-ray spectrum.
Therefore, the parameters of the fit were the Hydrogen column density ($N_\mathrm{H}$),
the plasma temperature (k$T$), the abundance (referred to the solar photosphere, $Z/Z_\odot$)
and the normalisation constant, which is directly related to the emission measure ($EM$).
Energy bins were grouped to contain, at least, 25 counts each.

A first attempt of fitting was done allowing all the parameters to vary. The results of
the fit for each observation are shown in Table 5 and Fig 3. Errors were determined with
XSPEC at a confidence level of 90 \%, following the recipes of the XSPEC 
manual\footnote{http://heasarc.nasa.gov/xanadu/xspec/xspec11/manual/manual.html}.
No clear differences in the temperature or global abundance are observed in the three 
orbital phases. However, NH vary notably at phase 0.48. The total X-ray flux also 
varies with the orbital phase being maximum at periastron and minimum at apastron.

It is well-known that, in the limit of low-counts, the column density and the global abundance
are degenerated for the fit(Arnaud 2004, Albacete-Colombo et al 2007) . 
Therefore, changing the value of any of them will affect the
result for the other. In order to check this effect in our data, we studied the space of parameters in the
model, using the command \textit{steppar} in XSPEC. For $N_\mathrm{H}$, we obtained that
its value is slightly correlated with $Z/Z_\odot$, while for temperature there is no such a correlation. Then, we decided
to fix the global abundance, since we see little or no change in its value during the three
phases covered by our data. We used $Z/Z_\odot = 0.26$, the value determined from the first two
observations. The temperature was also fixed to the mean value determined from the three
orbital phases (k$T = 0.55$ keV;  $T \sim 6.4\times 10^{6}$ K). Then, a new fit to the data was performed
for each observation. The results of the new fits are shown also in Table~5.
We see that the increase in $N_\mathrm{H}$ persists, while the little change detected in the
$EM$ for the third observation disappears when k$T$ and $Z/Z_\odot$ are fixed.

To investigate variations of the X-ray emission level of the source with time,
we obtained time-binned light curves by extracting the counts from the same regions than those used to extract the spectra
(for both source and background) in the energy range [0.3--10] keV.
Unfortunately, the count rates are very low and therefore no variability can be detected in short time periods
(of the order of few kiloseconds).
However, we found the mean count rate in this energy band to be variable and it varies as the
UV count rate measured with the OM (see Section~3.1). We note that
high-frequency (stochastic) events, i.e. flares, may be taking place during the observations. However,
the low count rate shown by the star prevents any detection of small flares. Large flares are much less
frequent and the probability of detecting one during short observations is very low. Nevertheless, the
low temperature detected during each observation does not suggest the presence of any kind of flaring
event during our observations.

\subsection{Analysis of the IUE data in low resolution}

The UV spectrum of AK~Sco is dominated by the photospheric contribution at long wavelengths, as expected in an 
F5 star (see Figure~5). At shorter wavelengths the spectrum is dominated by the tail of the
Balmer continuum and some prominent emission lines. In the near UV (2200$-$3200 \AA ), the dominant transitions
are the Fe~II and the Mg~II resonance multiplets. In the far UV (1200$-$1950 \AA), the most prominent
features are the resonance lines of H~I (Lyman $\alpha$), N~V[uv1], O~I[uv1], C~II[uv1], Si~IV[uv1] and 
C~IV[uv1] and the nebular, semiforbidden transitions of O~III] (1666 \AA), Si~III] (1892 \AA ) and
C~III] (1908 \AA ).  

IUE low resolution observations covered a broad range of phases. Spectra LWR14048 and SWP17804 were obtained 
at phase 0.97, spectra LWP12966 and SWP33197 at phase $\sim 0.02$ and spectra LWP 13009-13011 and SWP33241 
at phase $\sim 0.61$. The observations corresponding to the last two phases were obtained during the same 
cycle (see Table~4). The lines fluxes measured in the IUE spectra are given in Table~6. They have been extinction corrected using Valencic et al (2004) extinction law,  with A$_V$ and $R$ as in Table~1. The variation
of the lines flux is shown in Figure~6. Note that for rather similar phases (0.97 and 0.02) some lines
display significant variations (see also Table~6). However, for the observations obtained at different
phases (periastron and apastron) significant flux variations are detected in the O~I line, and to a smaller
degree in the the C~III] semiforbidden transition. These two lines are bona-fidae tracers of nebular (extended) 
emission in the TTSs environment (see e.g. G\'omez de Castro \& Verdugo, 2001 for the C~III] emission). 
Note also that the O~I line may be excited either by the Bowen mechanism from H~Ly~$\beta$ (Bowen 1947) or in the recombination of O~II. In both cases, the line excitation is expected to trace the end of the recombination cascade following photoionization by the stellar ionising radiation field, as suggested by GdCMA2012. In this context, the enhancement of the O~I flux at apastron would suggest a more efficient illumination of the circumbinary disk when the distance between both stars is the largest.

In Figure~7, the line fluxes have been normalised to the bolometric flux and
represented in the standard flux-flux diagrams for the TTSs atmospheric/magnetospheric diagnostics 
revised recently by GdCMA2012. 
AK~Sco is located on the TTSs regression line in the C~IV versus O~I diagram. It displays the characteristic excess emission of neutrals compared to high ionisation species. This excess is partially caused by the reprocessing of stellar
UV photons in the inner disk (GdCMA2012). However, AK~Sco, stands out of the TTSs regression line in the X-ray versus C~IV diagram. The source of this X-ray luminosity defect is unclear. The X-ray luminosities of the TTSs plotted in the diagram have been derived from the XEST survey (Guedel et al 2007), i.e. from XMM-Newton observations, as the AK~Sco data. The TTSs in the plot belong to spectral types from G0 to M5;  their X-ray/UV excess was found not to depend on the spectral type
(GdCMA 2012). For comparison, the X-ray/UV data of the main sequence stars have been plotted. Data have been extracted from Ayres et al (1995) for G to K types and from Linksy et al (1982) for M stars. The X-ray/UV excess of main sequence stars depends strongly on the spectral type with the largest excess being observed in the latest types. Ayres et al (1995) extended this study to F-type stars and showed that the correlation is broken at late F types; this is assumed to be caused by a decay in the dynamo activity since F-type stars represent an intermediate regime between stars with convective envelopes and stars with radiative envelopes. Typically, the X-ray luminosity  of F8-F0 main sequence stars is about an order of magnitude weaker than predicted by the main sequence regression line. AK~Sco data seem to extend this behaviour to PMS stars in a more dramatic manner: the X-ray flux drops by two orders of magnitude in spite of the strong tides. In this context we should remember that the X-ray fluxes of other PMS close binaries are also found to be
unexpectedly weak; DQ~Tau, UZ~Tau and KH~15D also seem to be subluminous in X-rays when compared with main sequence
stars of the same spectral types (Herbts \& Moran, 2005 and references therein). All the components in these systems are late type stars (K-M) and should not be affected by the transition between convective to radiative envelopes. There are however, PMS-SBs such as V826~Tau with normal X-ray luminosities (Carkner et al. 1996).  

\subsection{HST/STIS Spectra}

The high sensitivity of HST has allowed for the first time to obtain good SNR profiles of the main spectral lines
in the UV spectrum of AK~Sco. The profiles of the most prominent lines (Si~III (1206 \AA ), Ly$\alpha$, N~V[uv1], O~I[uv1], C~II[uv1], Si~IV[uv1], C~IV[uv1], O~III] (1666 \AA), Si~III (1892 \AA ), C~III (1908 \AA ) and Mg~II[uv1])
are displayed in Figure~8; all but the C~III] line have good  SNR in the HST observations. 

The profiles are very broad and asymmetric suggesting that line emission is produced in extended, unresolved structures. The Mg~II doublet is saturated since the 2796/2803 lines ratio is $\sim 1.4$ instead of 2 however, hot plasma tracers such as the Si~IV and C~IV [uv1] doublets have lines ratios close to their values in optically thin plasma. The blue edge velocity in the non saturated profiles is $\sim$350~km~s$^{-1}$. This value corresponds to the Keplerian velocity at $\sim 1.8$~R$_*$ which according to G\'omez de Castro (2009b) corresponds roughly to the extent of the emission region.

It is also noticeable the large broadening of the N~V line, with a Full Width Half Maximum (FWHM) of 278.7~km~s$^{-1}$. 
N~V is typically excited in the  transition region of the cool stars atmospheres at temperatures above some 40,000~K. 
As the thermal speed of the plasma at this temperature is 25~km~s$^{-1}$\footnote{The ion thermal velocity of fully ionized plasma at 40,000~K is
$v_{\rm th} = (k_B T_i/m_i)^{1/2}$, with $k_B$ the Boltzmann constant and $T_i$ and $m_i$, the ion temperature and mass, 
respectively. For $T_i = 40,000$K and $m_i = \mu m_H$,  $v_{\rm th} = 25$~km~s$^{-1}$, where the mean
molecular weight $\mu$ is 0.536, as corresponds to hot (fully ionized) plasma with solar abundances).}, 
the large broadening has to be caused by unresolved high velocity motions, confirming 
the existence of disturbed magnetospheres in AK~Sco stellar components (G\'omez de Castro 2009b).

The red wing of the profiles is weaker than the blue wing and seems to be partially absorbed in the resonance multiplets (Si~III, O~I, C~II, Si~IV, C~IV and Mg~II). However, the profiles corresponding to semiforbidden transitions of O~III] and Si~III] are also asymmetric suggesting an inherent asymmetry of the profiles. The asymmetry indicates that the warm matter close to the stars is  unevenly distributed in a high velocity field alike the expected, for instance, in spiral structures (see Sect.~4). 

The Ly$\alpha$ profile is completely different from the rest of the profiles since the blue wing is fully absorbed by the  Lyman H$_2$ band (Herczeg et al 2002) and also, presumably by a cool neutral outflow that it is not detected in the Mg~II line. The strongest H$_2$ transitions pumped by Ly$\alpha$ are detected in the spectrum
(see Figure~8 and Table~7).  The H$_2$ lines broadening is $\sim 30$~km~s$^{-1}$ that corresponds to a thermal velocity of about 10$^4$~K, characteristic of the photoevaporative flows from protostellar disks.

\subsection{Variations in the Mg~II profiles from IUE and HST}

The Mg~II profile varies significantly in the four observations available (see Figure~9) however, it does not seem to be a systematics when profiles corresponding to different cycles are compared (compare the profiles obtained at phase 0.56 and 0.64).  This fact, was already reported by A2003 for the H$\alpha$ line that it is excited in similar physical conditions than the Mg~II doublet; the H$\alpha$ profile at periastron is observed to be an inverse P-Cygni profile in one
cycle and a double peaked profile in another cycle. It is however, expected that the gas flow dragged from the inner border of the circumbinary disk contributes significantly to the formation of the lines and henceforth that the line profiles of
warm (few thousand Kelvins) plasma tracers, such as the Mg~II or the H$\alpha$ line, depend on the orbital phase. 

In this context, Hamilton et al (2012) have reported a correlation between the H$\alpha$ profile variations 
and the orbital phase in the KD~15D. In this system, the H$\alpha$ profile changes from an inverse P-Cygni during ingress
(as the stars approach the periastron) to a double peaked profile during egress (stars moves away from
periastron passage). Unfortunately, the Mg~II profiles of AK~Sco are few and obtained at singular points in the orbit, 
either at the periastron (phase 0.04) or very close to apastron (phases: 0.58, 0.64) and there is not any
clear trend between the profile asymmetry (the relative strength of the red to blue emission peaks)
and the orbital phase (see Figure~9).

\section{Numerical simulations of the binary-disk interaction}

Numerical simulations of the evolution of disks around pre-main sequence binaries show that the circumbinary disk is formed with an inner gap of radius about 2-3 times the semi-major axis of the orbit (Artymowicz \& Lubow 1994; Bate \& Bonnel 1997, Guenther \& Kley 2002, Hanawa et al 2010, Kaigorodov et al 2010, Fateeva et al 2011, de Val-Borro et al 2011, Shi et al 2012). In addition, stellar accretion disks may be formed around each component of the system depending on the flow rate through the inner disk gap to the stars and the properties of the binary system. To assist the interpretation of the X-ray, UV flux modulation and the broad non-thermal motions unveiled by the UV spectra, we have run a set of  numerical simulations of the interaction between the binary and the circumbinary disk. 

The details of the numerical scheme used in this paper are described in Bisikalo et al (2000) and Sytov et al (2011). The parameters of the system are listed in Table~1. The evolution of the system is parameterised in terms of the orbital phase as defined by the ephemeris, the rotation was assumed counter-clockwise  and the phase $\phi = 0$ corresponds to the periastron.  In this problem set-up, the positions of the components are defined by expressions

$$
\vec{r}_1=[(1-q')(\cos(\theta)-e)A,\quad(1-q')\sqrt{1-e^2}\sin(\theta)A,\quad 0]
$$
$$
\vec{r}_2=[-q'(\cos(\theta)-e)A,\quad-q'\sqrt{1-e^2}\sin(\theta)A,\quad 0]
$$

\noindent
where $q'=\frac{q}{q+1}$, and $q=\frac{M_1}{M_2}$ is the mass ratio of the components, $A$ is the semi-major axis, $\theta$ is the phase angle 
of the binary system and the phase $\phi$ is defined as $\phi={\theta}/{2\pi}$. For a given time ($t$) the angle $\theta$ is calculated from the implicit relation
$\theta=\frac{2\pi}{P_{orb}}t+e\sin(\theta)$.

The gas flow is described by the system of Euler equations:

\begin{center}
$\frac {\partial \rho}{\partial t} + \bigtriangledown \rho \vec{v}=0$\\
$\frac {\partial \rho \vec{v}}{\partial t} +\bigtriangledown (\rho \vec{v} \otimes \vec{v} ) + \vec{\bigtriangledown} P = -\rho \vec{\bigtriangledown} \Phi $\\
$\frac {\partial} {\partial t}( \rho (\varepsilon + \frac {v^2}{2}))
+\bigtriangledown \rho \vec{v}(\varepsilon+P/\rho + \frac {v^2}{2})=-\rho \vec{v} \vec{\bigtriangledown} \Phi$ , \\
\end{center}

\noindent where $\rho$ is the gas density, $\vec{v}$ is the velocity vector, $\varepsilon$ is the internal energy and $P$ is the gas pressure. An adiabatic equation of state is assumed with $P=(\gamma-1)\rho\varepsilon$ with $\gamma = 1.01$ to keep the solution close to the isothermal one. The variable gravitational potential, $\Phi$ is given by the following expression: 

$$
\Phi=-\frac{GM_1}{|\vec{r}-\vec{r}_1|}-\frac{GM_2}{|\vec{r}-\vec{r}_2|} 
$$
\noindent
The system of gas dynamics equations is solved numerically using the Roe-Osher-Einfeldt scheme (see Boyarchuk et al., 2002 and references therein).

\subsection{The disk model}

The circumbinary disk is considered as a Keplerian disk with constant equatorial density and hydrostatic vertical density profile:

$$
\rho(r,z)=\rho_d \exp{\left(-\left(\frac{|z|}{h(r)}\right)^{2}\right)},
$$

$$
h(r)=\frac {c_s}{\Omega(r)},\qquad\Omega(r)=\sqrt{G(M_1+M_2)/r^3},
$$
\noindent where $c_s$ is the sound speed, $\Omega(r)$ the angular velocity of the matter in the circum-binary dirk, $\rho_d$ the density in the orbital plane and $r$ the distance to the center of mass of the binary system. The temperature was assumed to be constant over the disk.

For all simulations we have run, the equatorial density and temperature of the disk were set to $\rho_{d} = 10^{-11} g/cm^3$ and $T_{d}=1000 K$, respectively.\\

The most important parameter of the numerical model is the value of the effective viscosity, $\alpha$. The effective viscosity (including contribution of the scheme, grid, waves, etc.) determines the rate of the angular momentum transfer in the disk, radial velocity $v_r$, and hence, the accretion rate, $\dot M_a$. The value of effective viscosity  in the simulations was equal to $\sim0.01$, that is in a good agreement with observations of accretion disks in cataclysmic variables. We use the boundary value of density $\rho_b$ to fit the numerical results to the observations, assuming that $\alpha$ (and $v_r$) is constant over the disk. Indeed, in the steady-state regime the accretion rate $\dot M_a$ equals the matter inflow through the outer boundary $\dot M$ thus, $\rho_b$ can be evaluated using the observed accretion rate.  Unfortunately, the accretion rate is not well determined from the observations of AK~Sco. The infrared spectral energy distribution can be properly fit only assuming reprocessing of the stellar radiation in the disk (A2003) setting up an upper limit to the accretion flow of $\sim 10^{-8}$~M$_{\odot}$~yr$^{-1}$. However, this value seems to be
too high for the observed UV excess of AK~Sco thus, a fiducial accretion rate of $10^{-9}$~M$_{\odot}$~yr$^{-1}$ has been used.

\subsection{The computational Set-Up}

The computational domain is a box of size: $12~A \times 12~A$ in the orbital plane (plane XY) by $\pm 0.75~A$ above/below it, in the Z-axis or vertical direction. The center of the Cartesian coordinate system is at the center of mass of the binary. The motion of the binary system in XY plane is set to be counter-clockwise. The domain is covered by an inhomogeneous grid with a total number of cells of $300\times 300\times 144$. In the inner region,  $1.5~A\times 1.5~A\times 0.2~A$, the cells have a constant size of $0.01A$. In the rest of the domain, the cell size increases to a maximum in the outer boundary of the computational domain as a geometric progression with factor 1.045. The outer boundary of the disk was defined by a cylinder with radius $R_{ext}=5.2A$ behind which the distribution of gas-dynamic values remains constant and defined by the initial conditions.

The initial conditions were adopted in the following manner: (1) the disk mass, velocity and temperature distribution as in Sect. 5.1; (2) to accelerate the simulations we assume the presence of the gap in the central region of the disk, i.e. the inner radius of the circumbinary disk was equal to 0.3~AU ($\simeq 2 \times A$), as derived from A1989 modelling of the infrared spectral energy distribution of AK~Sco.

The simulations have been left to run for $\sim 10$ orbital periods until quasi steady state is reached.

The inner boundary condition is set by the stars, represented by spherical surfaces of radii $2\times R_* \simeq 0.014~AU$. A free inflow condition is set at the boundaries of the cells intersecting the spheres at any time.  Cells released from the stellar surfaces are filled with ``vacuum'' values
$\rho_0$,$T_0$,$\vec{v}_0$, i.e., by highly rarefied gas. We adopted $\rho_0\approx10^{-20}g/cm^3$, $T_0\approx10^2 K$, $v_0=0$ for these cells.

\subsection{Modeling Results}

The structure of the gas flow obtained from the simulations consists of a circumbinary disk, a gap, circumstellar accretion disks, and a system of shock waves and tangential discontinuities. These elements are outlined in Figure~10. The spiral shock waves (bow shocks) formed due to the supersonic motion of the components of the binary system channel the gas within the gap into two flows, an outwards flow sending the material back to the inner border of the disk and an inwards flow that drags the gas into the stars, leading to mass accretion; these two flows are marked in Figure~10 
as I and II, respectively. Close to the stars, the mass flow enters into the Roche lobe; mass can then accrete,  either in the primary or in the secondary, marked as III and IV in the figure, respectively. A stationary shock wave forms in the area between both stars where the flows collide. 

Note that in the inner region of the circumbinary disk the velocity
distribution is non-Keplerian and gas motion is governed by bow shocks and
binary component’s gravity wells. We also found
that the size and the shape of the gap are substantially determined by the bow
shocks. The bow shock wave has a spiral shape, one end of the bow shock is
located near the circumstellar disk, while the other end resides inside the
circumbinary disk. In the flow structure this looks like if the bow shock
was connecting the stellar accretion disk to the circumbinary disk with a thick spiral
arm. There are two bow shocks in the system, one per star, making 
for a  two-arms spiral. However, those arms are not accretion streams and
they should not be misinterpreted as such flow features inside which the
matter flows from the circumbinary disk to the accretion disk. When the
matter of the gap passes through the bow shock it forms the dense spiral
behind the wave, but some part of it flows to the accretion disk and the
rest of it flows back to the circumbinary disk. These two flows are
separated by the head-on collision point visible in the velocity field on
the forefront of the bow shock. Remarkably, these streams are not
distinguishable on the density distribution as two separate streams, indeed,
they look like a single thick spiral arm.

Figure~11 displays the evolution of the inner region of computational domain during a single orbital period. Twelve panels show the density distribution at different times (phases) evenly distributed over the orbital period, starting at periastron. The AK Sco binary system has a substantially elliptical orbit which results in an interesting phenomenon. When the system approaches periastron, the outer boundaries of the circumstellar disks (and the accretion streams passing by) get close enough one to each other to effectively lose the angular momentum, leading to an increase of the accretion rate by a factor of 2-3 (see Figure 12). A tangential discontinuity is produced at the point where the primary and secondary accretion flows get in contact because, at the interface between them, the matter orbiting around the primary moves in the opposite direction that the matter orbiting the secondary (see 
Figures 11 and 13). Collisional heating, development of instabilities (like Rayleigh-Taylor instability on the interface), shock wave formation will efficiently remove the angular momentum and, henceforth, lead to an accretion outburst.

The structure of the gas flow in the orbital plane of the system at periastron and apastron is shown in more detail in Figures~13 and 14, respectively. The velocity field presented on Figs. 13, 14 shows the direction of all gas flows in the system. The velocity distribution along with shock waves 
(shown in these figures as density enhancements) give a full description of the flow structure in the
system. At periastron passage, a complex pattern of shock waves appears between the components of the binary system.
Since the accretion rate significantly increases at periastron this results in 
an increase of the high energy radiation flux. At the same time the gap becomes cleaner, i.e. contains less matter, at periastron than at apastron mainly due to the significant increasing of the accretion rate and more efficient removal of the matter from the gap, by stronger shock waves caused by faster orbital motion. As a result, the total amount of mass in the gap is seen to vary, as expected, in anti-phase with the accretion rate.

\section{A general scenario for AK~Sco}

Based on the three XMM-Newton observations that scarcely monitor the system from periastron to apastron, there seems to be an orbital modulation of the high energy tracers of magnetospheric activity. The energy radiated in these tracers, both in the UV and X-ray, increases with the  tide and the accretion rate: it is higher at  periastron than at apastron. Moreover, the maximum flux is observed at phase $\phi =0.117$, when stellar occultation could occur (A1989), thus precluding any association with eclipse like phenomena.

In Figure~15, the variation of the radiative energy release during the binary cycle is shown for X-ray, UV 
wavelengths (from this work) in the bottom panels. At the top, data from the optical monitoring of AK~Sco run 
in 1982 by Manset et al (2005, hereafter MBB2005) are displayed for comparison. The variation of the U and V magnitude, as well
as the variation of the polarisation in the U-band during the orbital cycle are plotted. Note that 
U and V magnitudes remain around a basal level between periastron and phase 0.4; after the flux increases
significantly to reach the maximum close to apastron. Finally, at phase 0.65, the flux decreases rapidly at 
the same time that the polarisation increases. The correlation between the increase of polarisation and the
decrease in the optical bands flux is interpreted by MBB2005 as a dusty structure crossing the line of sight.

Returning back to our observations, X-ray data clearly show a variation of the gas column during
the orbit. The gas column varies from $N_H = 0.35^{+0.29}_{-0.29} \times 10^{21}$~cm$^{-2}$ at periastron to 
$N_H = 1.11^{+0.49}_{-0.46} \times 10^{21}$~cm$^{-2}$ at apastron (see column 3 in Table 5). This variation can be translated into an $A_V$ variation making use of the conversion from Hydrogen column density to optical extinction derived recently by Guever \& Oezel (2009) from X-ray based H~I column densities in the Galaxy: $N_H = (2.21 \pm 0.09) \times 10^{21} A_V$(mag). According to this conversion, $A_V$ would vary from  0.16 mag at periastron to 0.50 mag at apastron; this last value agrees with the fiducial A$_V$ value determined from optical observations (see Table~1). As it will be shown below, this variation in gas column is consistent with the variation in 
polarisation detected by MBB2005.

Serkowski et al. (1975) established that polarisation by interestellar dust takes its maximum value at a wavelength of $ \lambda_{max} \simeq 0.45 \mu m - 0.8 \mu m$.  This maximum polarisation ($p_{max}$) correlates with
with E(B-V) as, $ p_{max} \leq 9.0 E(B-V)$ (Whittet et al 1992). 
Since $R=4.3 = A_V/E(B-V)$ in AK~Sco (MBB2005), this expression can be transformed into
$ p_{max} \leq 2.1 A_V = 0.95 N_{H,21}$~cm$^{-2}$, with $N_{H,21} = N_H/10^{21}$ making use again of Guever \&
Oezel (2009) conversion. According to this scaling, a variation in $N_H$ as the derived from the XMM/EPIC observations would correspond to a variation in $p_{max}$ of 0.7\%, in good agreement with the polarisation measurements (see Figure~15). 
Note that for the large sample of stars studied by Serkowski et al (1975), the median value is $<\lambda_{max}> = 0.55 \mu$m
which roughly corresponds to the V-band and agree well with MBB2005. 

In summary, the maximum gas column detected in the XMM-Newton observations, in 2011, is comparable to the maximum gas/dust column reported by MBB2005 from a photometric/polarimetric optical monitoring of the system in 1982. This suggests that the absorption could be produced by a rather estable pattern, such as warp waves in the disk, instead of tracing a casual passage of a dust cloud by the line of sight. Yet, further observations are required to confirm this point.

The variable extinction affects significantly the UV light curve obtained with OM. A variation, 
$\Delta A_V$, of 0.34 mag from periastron to apastron, corresponds to an increment in $\Delta A_{UV}$ in the OM band of 0.82 mag. This implies that while the X-ray flux would be enhanced at periastron, the UV flux 
in the OM-UVM2 window would be maximum at apastron. This, apparently puzzling, behaviour is consistent
with the detection of an enhancement of the O~I flux (see Sect.~3.3). Note that the UV flux detected
by OM comes from two main sources:
\begin{enumerate}
\item The {\it direct} flux from the stellar magnetospheres and the accretion shocks, that it is expected to be maximum at periastron since the accretion rate (and the X-ray flux) is maximum at this phase (see also Fig.~ 12).
\item The high energy (X-ray, extreme and far UV) stellar flux {\it reprocessed} in the circumstellar environment and diffused into the line of sight. 
\end{enumerate}

Since the OM-UVM2 flux is higher at apastron than at periastron, {\it reprocessed}
radiation must be the dominant contributor to the UV flux in this band. This is consistent with the
results from the simulations shown in Figure~11. The variable gravitational field induced by the binary in the 
disk produces a variable distribution of circumstellar material; it acts like a gravitational piston that
drags material from the inner border of the disk into the gap. At periastron, there is an accretion
outburst and the gap becomes rather clean thus, though the {\it direct} UV flux is higher, the contribution
from the {\it reprocessed} UV radiation is minimum. At apastron however, the gap is filled with gas and the stars
are closer to the inner border of the circumbinary disk than at periastron.  Henceforth, the 
reprocessing of the high energy radiation from the star into lower energies, as well as the diffusion
of this radiation into the line of sight is more efficient. This provides a general scenario to interpret
AK~Sco data that it is also consistent with previous results in the optical range; 
note that as the {\it variation} in the optical bands 
is also dominated by {\it reprocessed} radiation (the Balmer continuum in the U band and
the emission lines in the V band), the flux should also be maximum at apastron in these bands.

\section{Conclusions}

This work has allowed us to define a working scenario to explain the main properties of AK~Sco. This star has revealed itself to be a unique new type of interacting binary system. We have shown that AK~Sco eccentric orbit acts like a gravitational piston. At apastron, matter is  efficiently dragged from the inner disk border, filling the inner gap and producing accretion streams that end as ring-like structures around each component of the system. At periastron, the ring-like structures get in contact, leading to angular momentum loss, and thus producing an accretion outburst. Additional relevant conclusions from this work are:

\begin{itemize}

\item In spite of its peculiarities, AK~Sco satisfies the same UV flux-flux relations than the rest of the TTSs. Thus, 
the strong tidal wave, driven by the close and highly eccentric orbit, does not seem to modify the distribution of the 
energy released among the various UV spectral tracers.  

\item AK~Sco X-ray luminosity is two orders of magnitude smaller than the one observed in  the TTSs. It is unclear whether this is mimicing the well known drop of the magnetic activity around F8 types in main sequence stars or whether it could be associated with other physical processes since not all UV bright and accreting TTSs are detected in X-ray.

\item The modest monitoring run with XMM-Newton suggests that accretion is enhanced at periastron in anti-phase with the filling of the inner gap of the disk, and reprocessing of high energy radiation into the UV recombination bands.  This behaviour agrees with the predictions from numerical simulations.

\end{itemize}

\acknowledgments

This work has been partly funded by the Ministry of Science and
Innovation of Spain through grant, AYA2008-06423-C03-01/03. AS and DB were 
partly supported by RFBR and Russian Federation President grants.

{\it Facilities:} \facility{XMM-Newton (OM)}, \facility{XMM-Newton (EPIC)}.


\newpage
\begin{figure}[h]
   \includegraphics[width=0.80\textwidth]{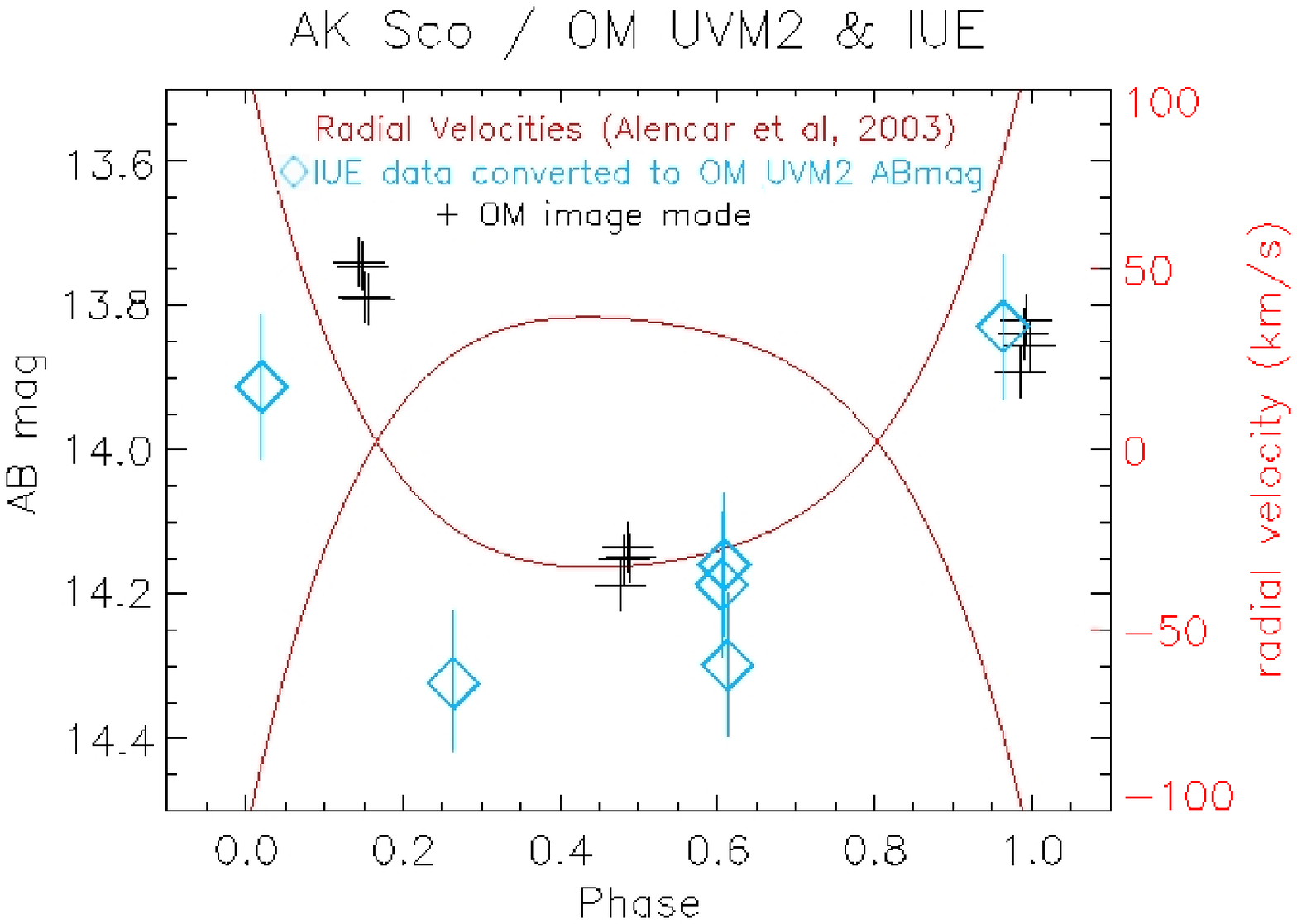}
 \caption{Observations of AK Sco with IUE and OM. AB magnitudes are overploted on the radial velocity curve as derived 
by A2003. Error bars are plotted for  the IUE based measurements. Error bars are negligible in this plot for OM observations.
}
\label{fig:lc}
\end{figure}

\newpage
\begin{figure}[h]
   	\includegraphics[width=0.80\textwidth]{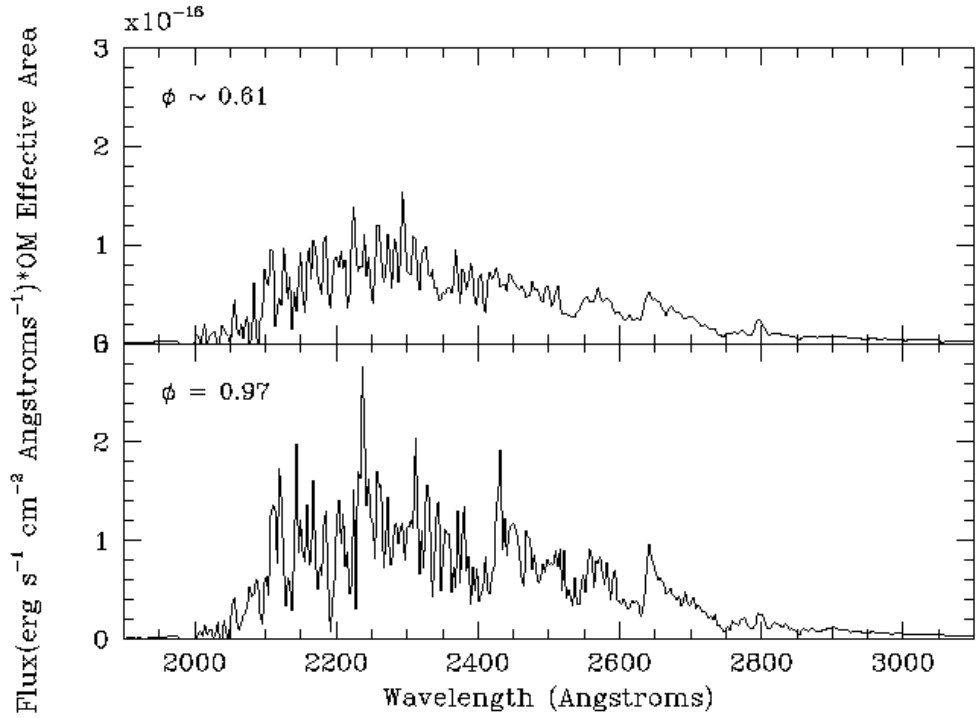}
 \caption{IUE UV spectrum of AK~Sco convolved with OM UVM2 response at two orbital phases.
The flux is dominated by the resonance transitions of  Fe~II(uv1) (2585-2632 \AA ),
Fe~II(uv2) (2360-2410 \AA ) and Fe~II(uv3) (2327-2380 \AA ) and the tail of the Balmer continuum.
is }
\label{fig:iueom}
\end{figure}

\newpage
\begin{figure}[h]
   \includegraphics[width=0.80\textwidth]{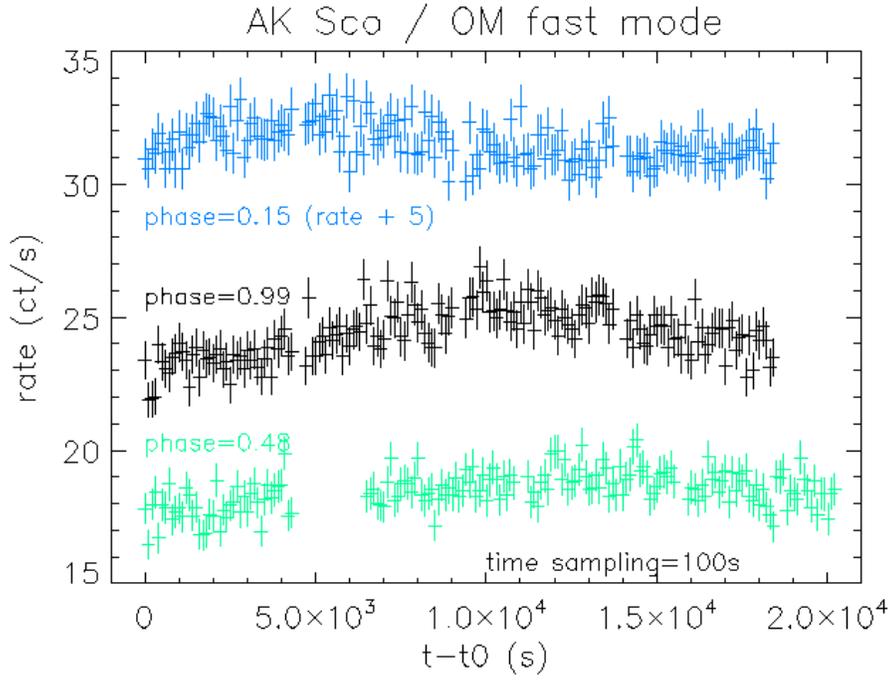}
 \caption{Light curves of AK Sco obtained with OM in fast mode.The light curve
corresponding to phase 0.15 has been displaced by 5 cts/s upward for more
clarity.}
\label{fig:om2}
\end{figure}

\newpage

\begin{figure}[h]
\begin{center}
 	\includegraphics[width=0.80\textwidth]{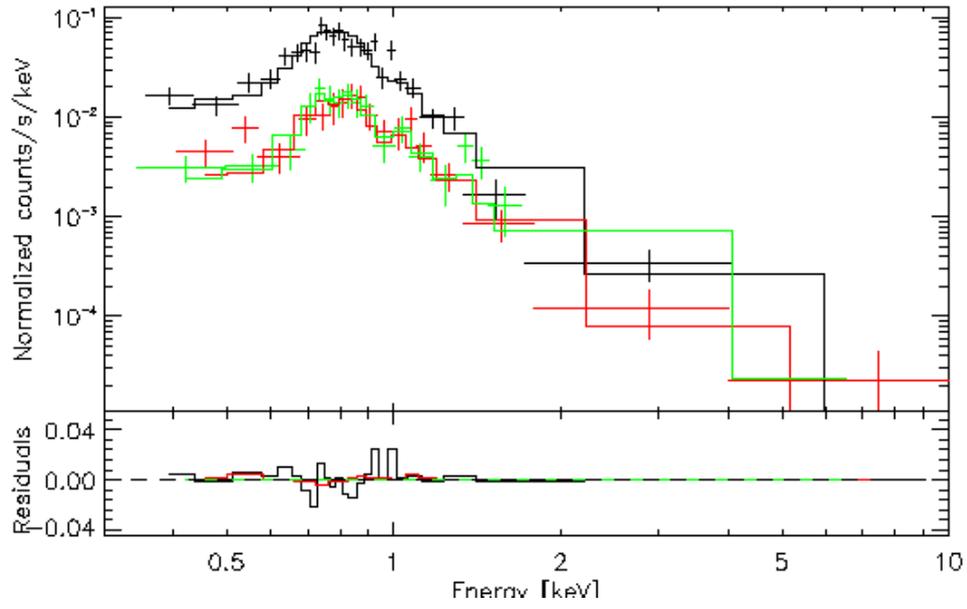}\\
  \end{center}
\caption{Data and folded model for EPIC observation ID.0654140201 of AK~Sco.
Data are marked with crosses and models with lines. Colours are used to
code EPIC-PN data (black), EPIC-MOS1 (red) and EPIC-MOS2 (green).
}
\label{fig:XraySED}
\end{figure}

\newpage
\begin{figure}[h]
   	\includegraphics[width=0.80\textwidth]{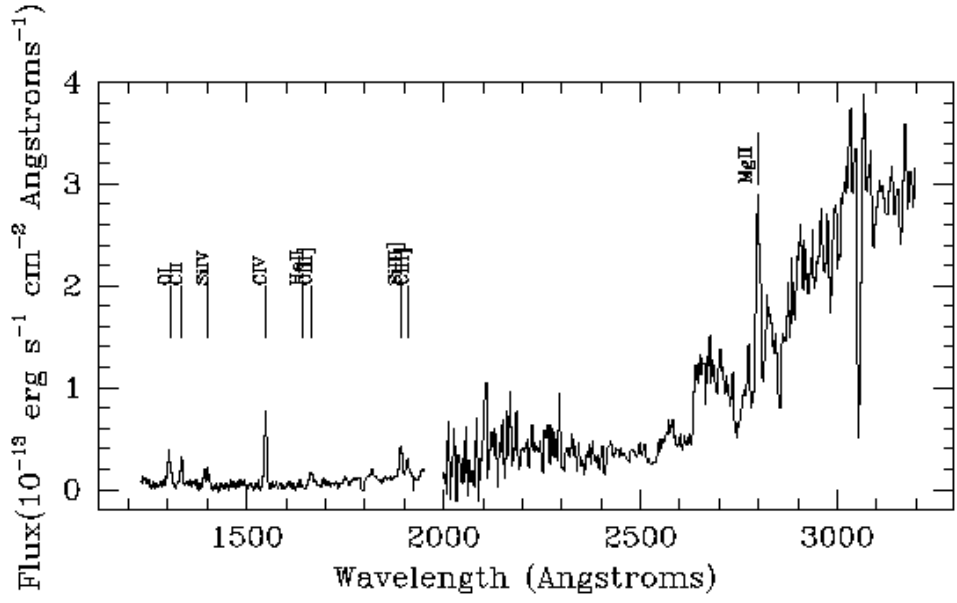}
 \caption{The low dispersion IUE spectrum of AK~Sco. {\bf MODIFY LABELS!!}}
\label{fig:IUEAKsco}
\end{figure}

\newpage
\begin{figure}[h]
   	\includegraphics[width=0.80\textwidth]{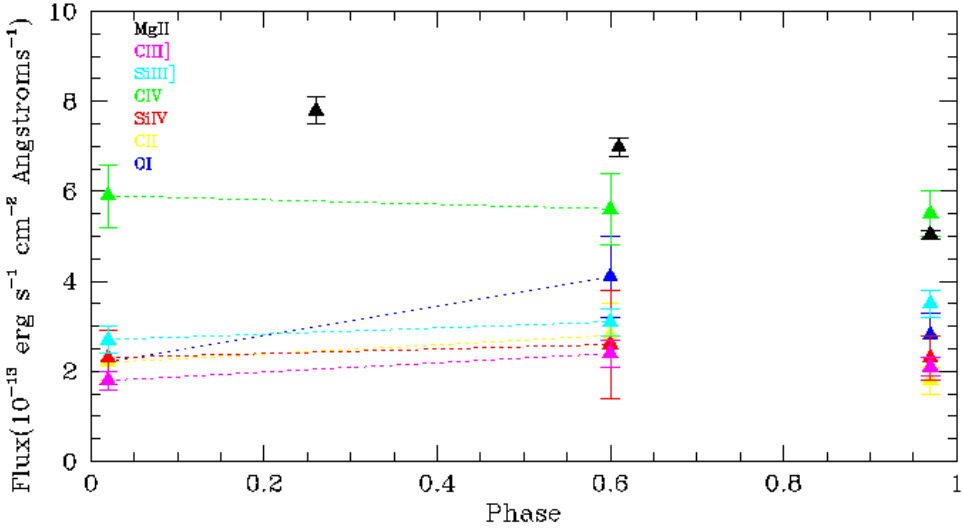}
 \caption{Variation of the fluxes of the main spectral lines in the UV spectrum of
AK Sco, from low dispersion IUE data. 
}
\label{fig:linesflux}
\end{figure}

\newpage
\begin{figure}[h]
\begin{center}
   	\includegraphics[width=0.5\textwidth]{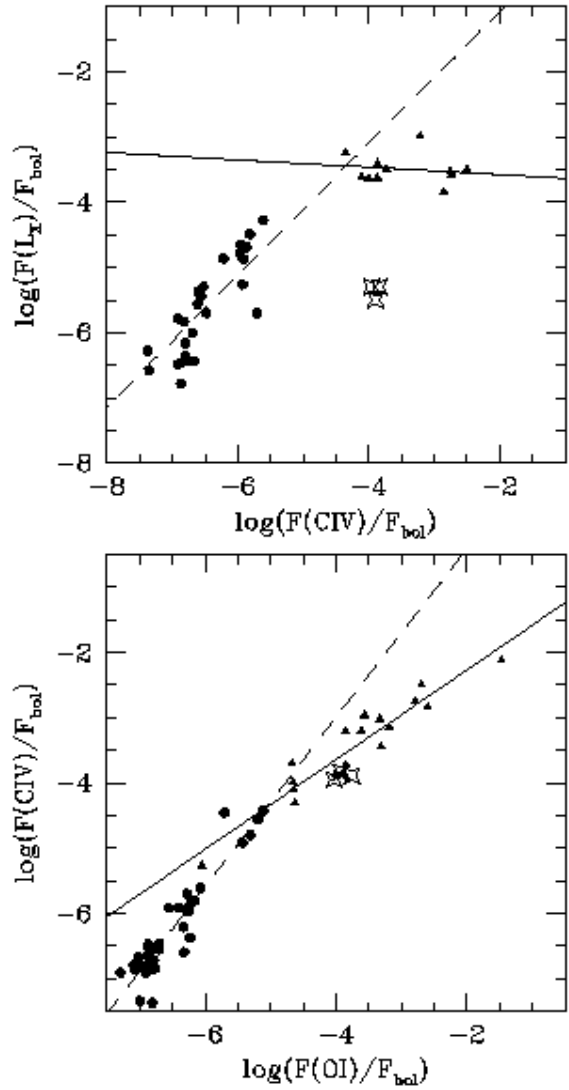}
\end{center}
 \caption{Location of AK~Sco in the flux-flux diagrams of cool main sequence
stars and TTSs as in G\'omez de Castro \& Marcos (2012). {\it Top:} X-ray versus
C~IV; {\it Bottom:} C~IV versus O~I. The filled circles represent the
location of late type main sequence stars in the diagram from Ayres et al (1995)
and Linsky et al (1982).
The filled triangles represent the location of TTSs from G\'omez de Castro \& 
Marcos-Arenal (2012). The regression lines for late type stars and TTSs are represented by
dashed and solid lines, respectively. The location of AK~Sco is marked with stars.
}
\label{fig:colour-colour}
\end{figure}

\newpage

\begin{figure}[h]
   	\includegraphics[width=14cm]{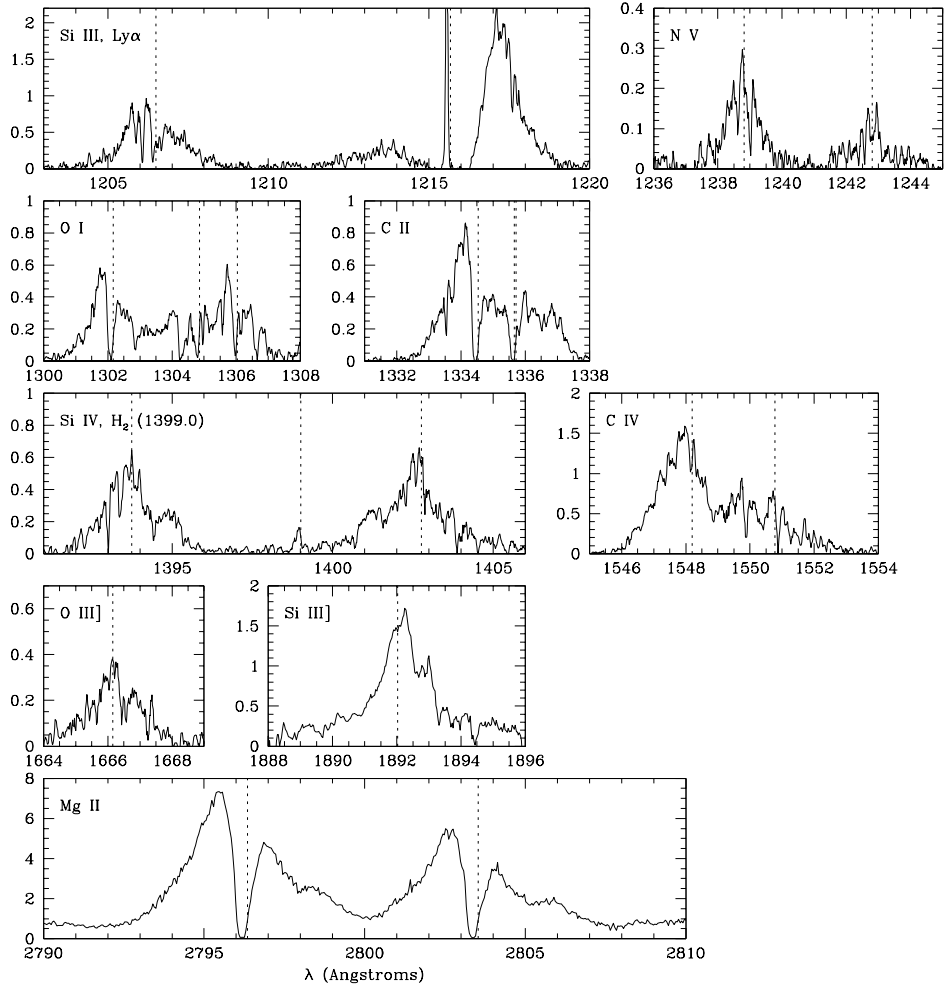}
 \caption{Profiles of the main spectral lines observed in the UV spectrum of
AK~Sco with the HST/STIS. The rest wavelength of the transitions is marked by dashed lines.
Fluxes are in units of $10^{-13}$ erg~s$^{-1}$~cm$^{-2}$\AA $^{-1}$. }
\label{fig:hstprof}
\end{figure}

\newpage

\begin{figure}[h]
   	\includegraphics[width=0.80\textwidth]{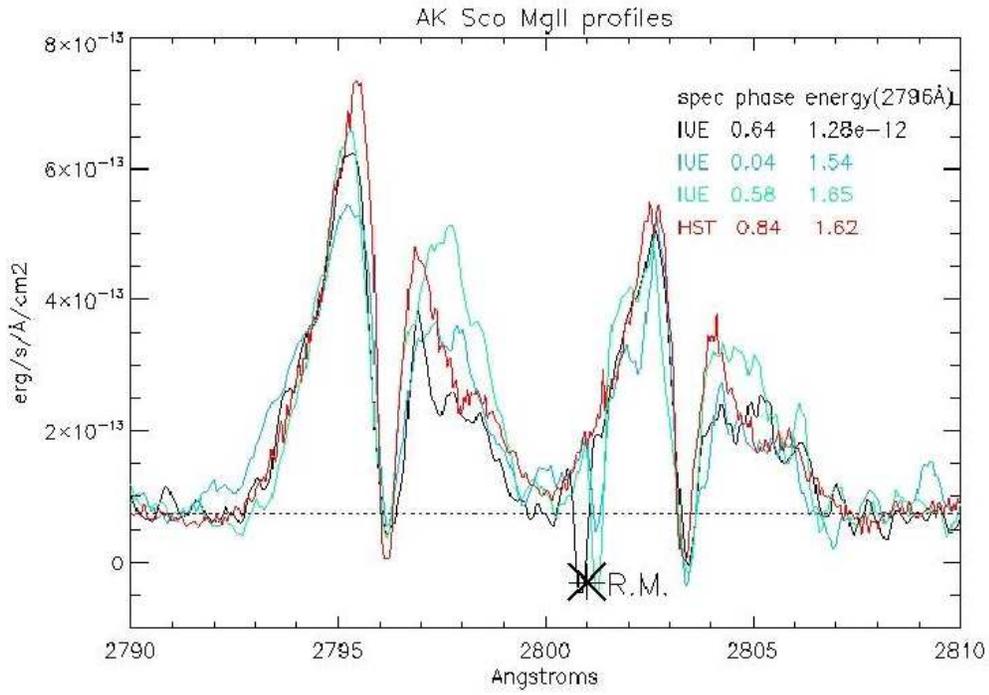}
 \caption{Variability of the Mg~II profiles of AK~Sco from the high dispersion spectra in the IUE and HST Archives. RM denotes a IUE fiducial reseau mark contaminating the spectrum at $\sim 2801 \AA$.}
\label{fig:mg2}
\end{figure}

\newpage

\begin{figure}[h]
\centering
\includegraphics[width=15cm]{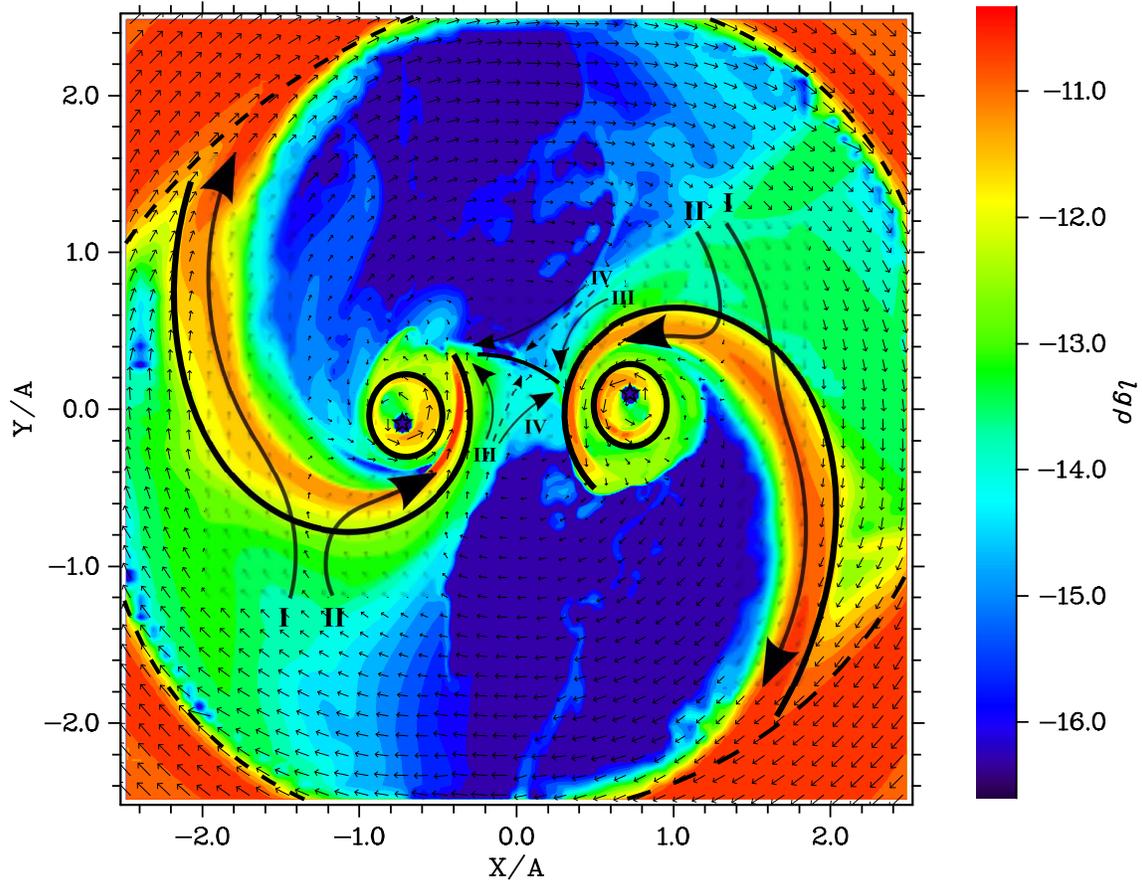}
\caption{The ﬁgure shows the location of the system components,
accretion disks with corresponding bow shocks, the
stationary shock wave between accretion disks and few
major gas ﬂows: (I) - the outﬂow to the circumbinary
envelope; (II) - the accretion stream; (III, IV) - parts
of accretion streams that form the inner stationary shock
wave and contribute to the accretion rates. The dashed line indicates
the boundary of the gap.
}
\label{fig:sketch}
\end{figure}

\newpage

\begin{figure}[h]
\centering
\includegraphics[width=15cm]{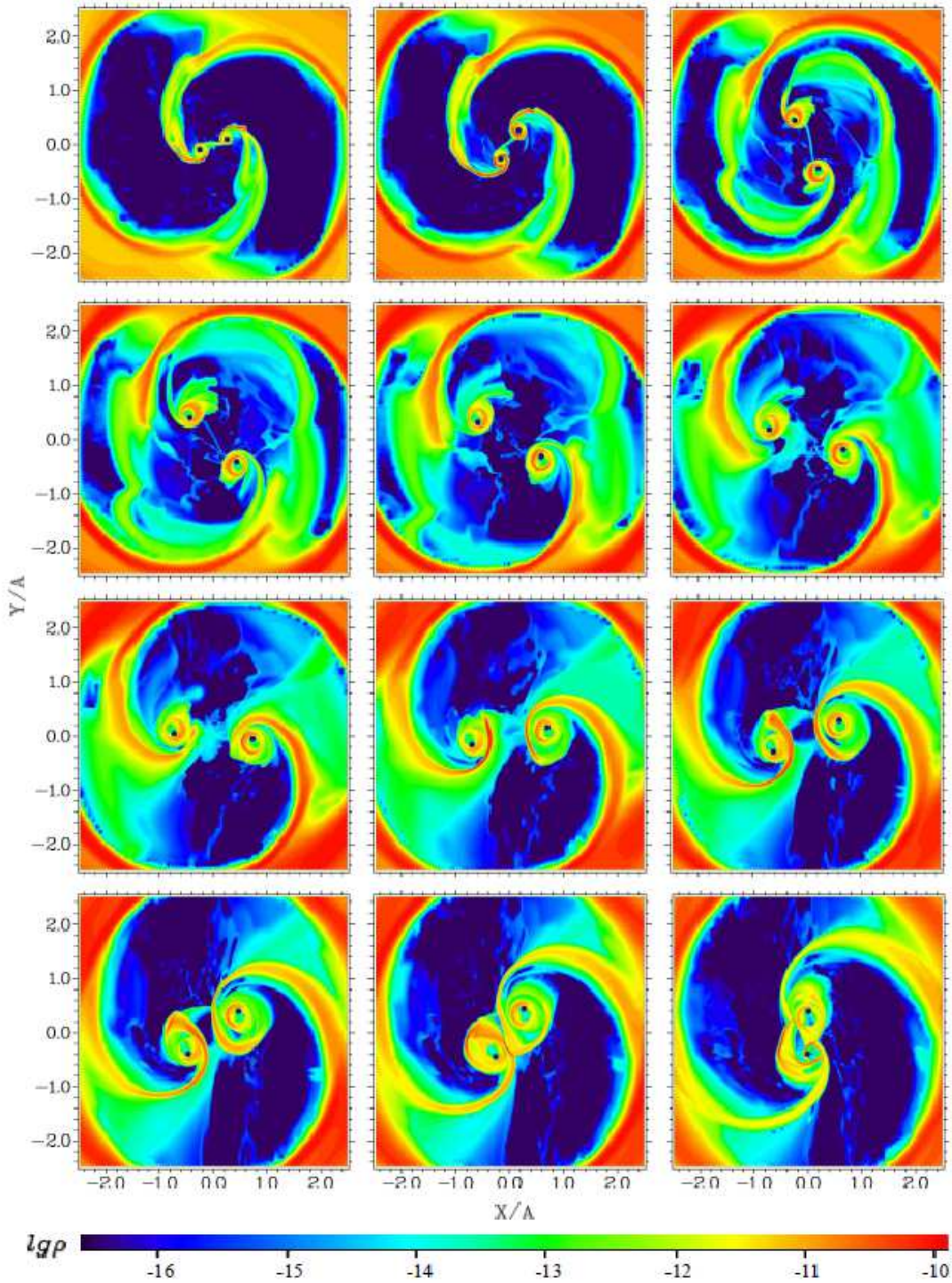}
\caption{Evolution of the density distribution in the inner gap with the 
orbital period. Starts at the upper, left corner at $\phi = 0$, at 
periastron, with an incremental phase of $\Delta \phi = 0.083$ from frame to 
frame running to the right in the rows and, then to the bottom, row by row. 
Each frame represents a physical size of 5A by 5A. }
\label{fig:mosaic}
\end{figure}

\newpage
\begin{figure}[h]
\centering
\includegraphics[width=15cm]{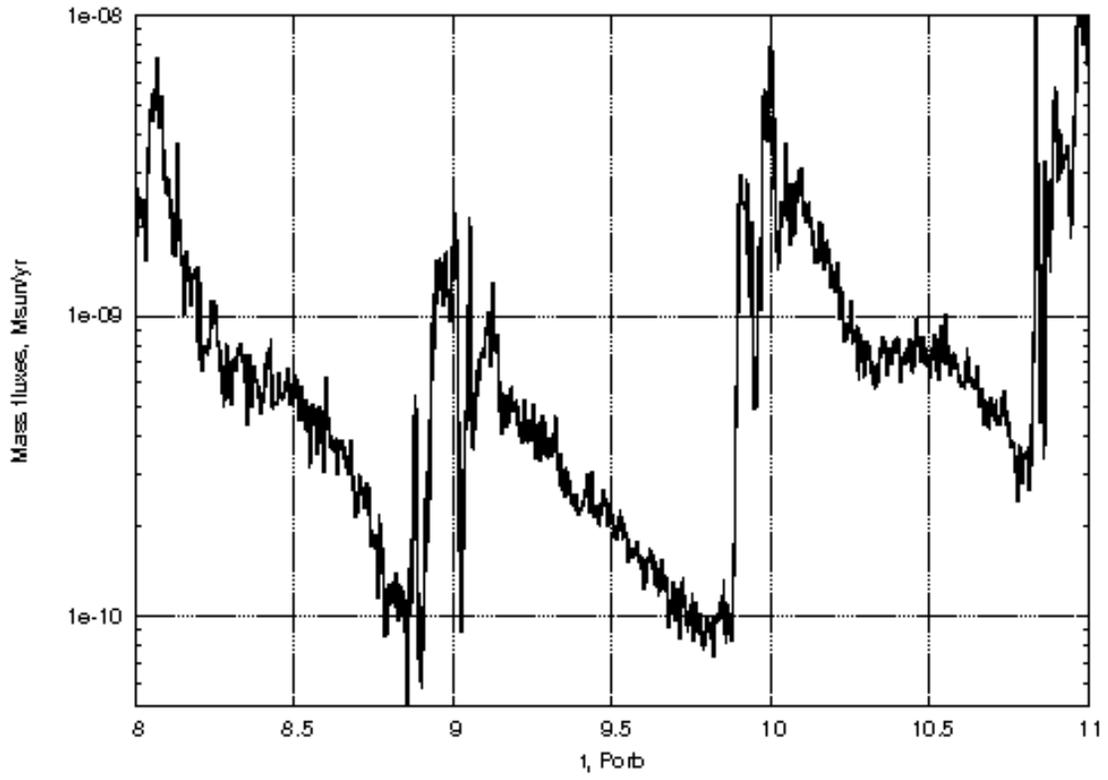}
\caption{Accretion rate on the primary. Notice that accretion
outbursts occur at periastron passage.
}

\end{figure}

\newpage
\begin{figure}[h]
\centering
\includegraphics[width=15cm]{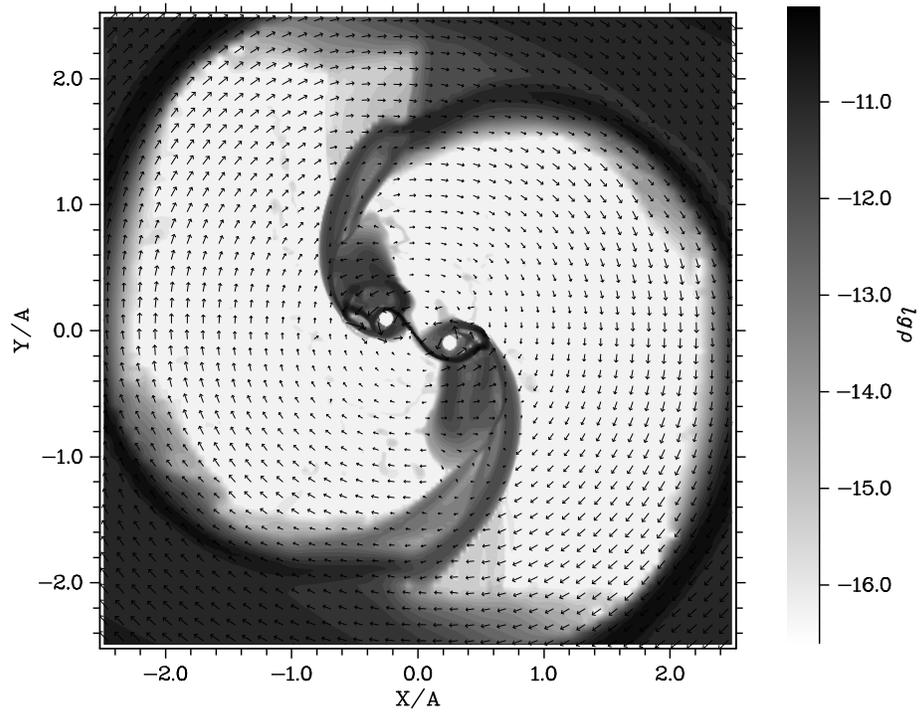}
\caption{Density and velocity distribution in the orbital plane of the binary system at periastron passage. The density is displayed in the laboratory reference frame, but velocity vectors are translated to the rotating stellar frame 
to assist the interpretation. }\label{fig:periastron}
\end{figure}

\newpage
\begin{figure}[h]
\centering
\includegraphics[width=15cm]{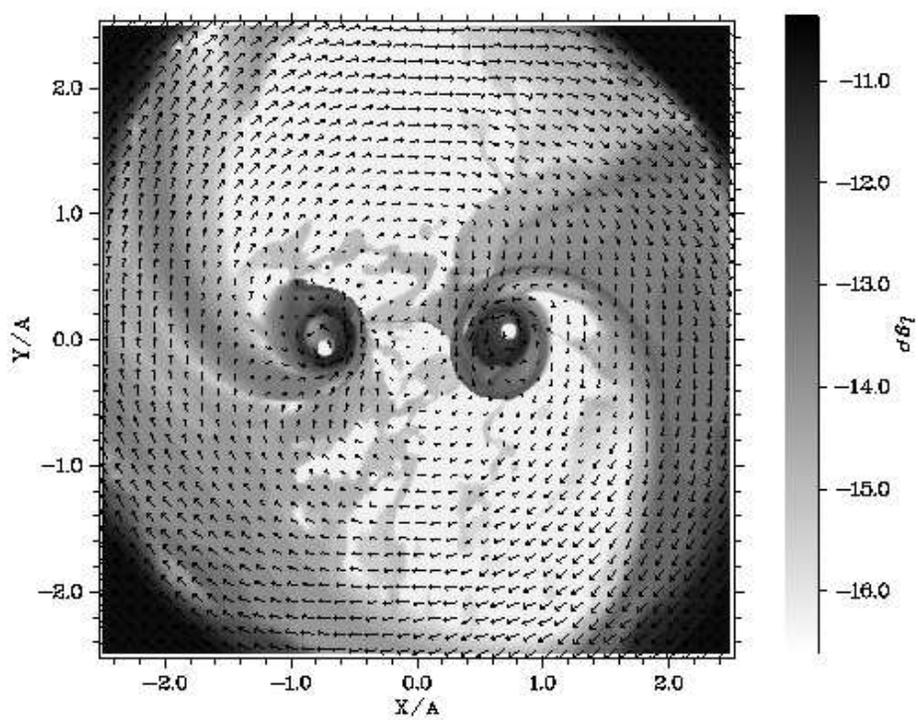}
\caption{As Figure 13  at apastron passage.  }\label{fig:apastron}
\end{figure}

\newpage
\begin{figure}[h]
   	\includegraphics[width=0.80\textwidth]{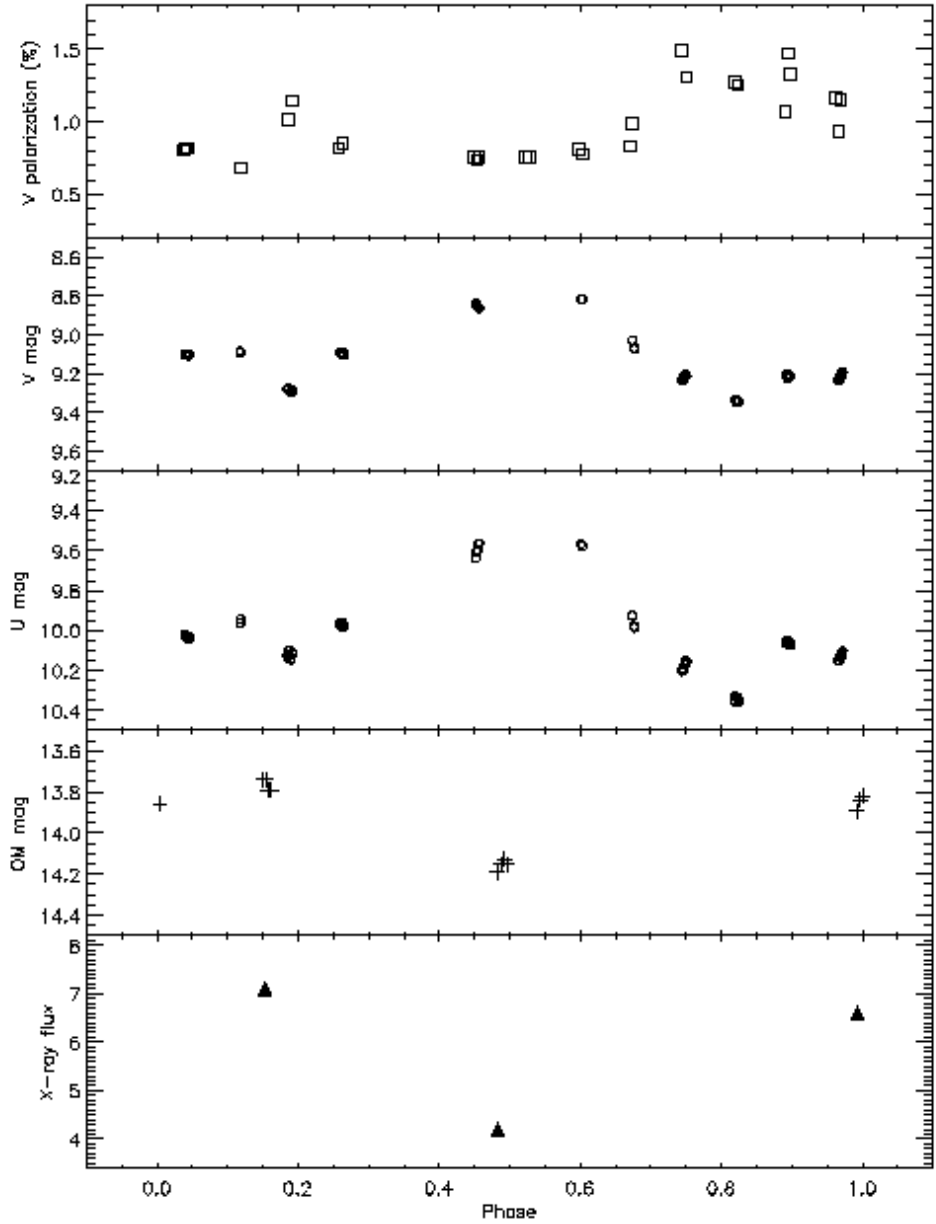}
 \caption{Summary of AK~Sco variability during the orbital cycle. Optical polarisation (V-band), 
U and V magnitudes are plotted in the three upper panels using data from MBB2005. 
UV and X-ray variations are plotted from data in this work.}

\label{fig:summary}
\end{figure}


\newpage
\begin{table*}
\caption{AK Sco main parameters.}
\begin{center}
\small
\begin{tabular}{lll}
\hline
Property & Value & Source \\
\hline
Projected semimajor axis & $a\sin i = 30.77 \pm 0.12$R$_{\odot}$ & Andersen et al 1989 \\
Eccentricity & e= 0.47 & Andersen et al 1989, Alencar et al. 2003 \\
Orbital period & P=13.609 d & Andersen et al 1989, Alencar et al. 2003 \\
Inclination & $ i=65^o-70^o$ & Alencar et al. 2003 \\
Age & 10-30 Myrs & Alencar et al. 2003  \\
Spectral type & F5 & Alencar et al. 2003\\
Stellar Mass & $M_* = 1.35 \pm 0.07$M$_{\odot}$ & Alencar et al. 2003\\
Radius & $R_* = (1.59 \pm 0.35) R_{\odot}$ & Alencar et al. 2003 \\
Projected rotation velocity & $v \sin i = 18.5 \pm 1.0$ km s$^{-1}$ & Alencar et al. 2003 \\
Bolometric flux & $6.33\times 10^{-9}$ erg~s$^{-1}$cm$^{-2}$ & Andersen et al 1989\\
Extinction: A$_V$ & 0.5 mag & Manset et al. 2005 \\
Extinction: R & 4.3 & Manset et al. 2005 \\
Distance & 102.8 pc & Van Leeuwen 2007 \\
\hline
\end{tabular}
\end{center}
\label{tab:table1}
\end{table*}

\newpage

\begin{table*}
\caption{XMM-Newton observations of AK Sco}
\begin{center}
\small
\begin{tabular}{l c c c l l}
\hline
 Observation  & Instrument  & Mode & Filter & Exposure start time & Duration (s)\\
\hline
0654140201 & EPIC-pn   & Full frame & Thick &  2011-03-15T21:26:40  & 24533 \\
           & EPIC-MOS1 & Full frame & Thick &  2011-03-15T21:04:19  & 26119 \\
           & EPIC-MOS2 & Full frame & Thick &  2011-03-15T21:04:18  & 26125 \\
           & RGS-1     & Spectro + Q&       &  2011-03-15T21:03:32  & 26413 \\
           & RGS-2     & Spectro + Q&       &  2011-03-15T21:03:42  & 26407 \\
           & OM        & User def.  & UVM2  &  2011-03-15T21:12:38  & 4400 \\
           & OM        & User def.  & UVM2  &  2011-03-15T22:31:04  & 4400 \\
           & OM        & User def.  & UVM2  &  2011-03-15T23:49:32  & 4400 \\
           & OM        & User def.  & UVM2  &  2011-03-16T01:08:00  & 4400 \\
\hline								
0654140301 & EPIC-pn   & Full frame & Thick &  2011-03-18T01:04:36  & 23833 \\
           & EPIC-MOS1 & Full frame & Thick &  2011-03-18T00:42:17  & 25417 \\
           & EPIC-MOS2 & Full frame & Thick &  2011-03-18T00:42:14  & 25425 \\
           & RGS-1     & Spectro + Q&       &  2011-03-18T00:41:28  & 25714 \\
           & RGS-2     & Spectro + Q&       &  2011-03-18T00:41:36  & 25706 \\
           & OM        & User def.  & UVM2  &  2011-03-18T00:50:33  & 4400 \\
           & OM        & User def.  & UVM2  &  2011-03-18T02:09:01  & 4400 \\
           & OM        & User def.  & UVM2  &  2011-03-18T03:27:29  & 4400 \\
           & OM        & User def.  & UVM2  &  2011-03-18T04:45:54  & 4400 \\
\hline								
0654140401 & EPIC-pn   & Full frame & Thick &  2011-03-22T13:39:54  & 27034 \\
           & EPIC-MOS1 & Full frame & Thick &  2011-03-22T13:17:35  & 28618 \\
           & EPIC-MOS2 & Full frame & Thick &  2011-03-22T13:17:33  & 28625 \\
           & RGS-1     & Spectro + Q&       &  2011-03-22T13:16:48  & 28914 \\
           & RGS-2     & Spectro + Q&       &  2011-03-22T13:16:56  & 28909 \\
           & OM        & User def.  & UVM2  &  2011-03-22T13:25:53  & 4400 \\
           & OM        & User def.  & UVM2  &  2011-03-22T15:14:20  & 4400 \\
           & OM        & User def.  & UVM2  &  2011-03-22T16:32:46  & 4400 \\
           & OM        & User def.  & UVM2  &  2011-03-22T17:51:14  & 4400 \\
\hline
\end{tabular}
\end{center}
\label{tab:table2}
\end{table*}

\newpage
\begin{table}

\caption{Photometry of AK Sco with OM using the UVM2 filter (2310 \AA)}


\begin{tabular}{l l l l l}
\hline

 Observation  & HJD      & phase & AB mag  & Flux  \\
           &  -2400000.5 &        &     & (erg~s$^{-1}$cm$^{-2}$\AA $^{-1}$) \\
\hline
0654140201 &    55635.91003  & 0.9919   & 13.89   & 5.62e-14  \\                   
           &    55635.96450  & 0.9959   & 13.84   & 5.91e-14  \\
           &    55636.01900  & 0.9999   & 13.82   & 6.01e-14  \\	                   
           &    55636.07350  & 0.0039   & 13.86   & 5.82e-14  \\
\hline	      
0654140301 &    55638.06157  & 0.1500   & 13.74   & 6.48e-14  \\           
           &    55638.11606  & 0.1540   & 13.74   & 6.45e-14  \\	                   
           &    55638.17056  & 0.1580   & 13.79   & 6.19e-14  \\	                  
           &    55638.22502  & 0.1620   & 13.79   & 6.17e-14  \\
\hline	      
0654140401 &    55642.58653 & 0.4824   & 14.19   & 4.28e-14  \\	                   
           &    55642.66185 & 0.4879   & 14.15   & 4.43e-14  \\	                   
           &    55642.71633 & 0.4919   & 14.13   & 4.50e-14  \\	                   
           &    55642.77082 & 0.4960   & 14.15   & 4.44e-14  \\
\hline
\end{tabular}
\label{tab:table3}
\end{table}

\newpage
\begin{table*}
\begin{center}
\caption{UV observations of AK~Sco.\label{tab:table4}}
\begin{tabular}{lllll}
\hline\hline
\multicolumn{5}{c}{IUE Observations} \\
\hline
 Observation  & HJD         & Texp  & phase & FES V-mag   \\
              & -2400000.5    & (min) &       &           \\
\hline
LOW DISPERSION & & & & \\
\hline
 LWR14048  &   45210.77531  & 12  &   0.96  & 8.84 \\
 SWP17804  &   45210.85330  & 170 &   0.97  & 8.84 \\
 SWP33197  &   47252.82195  & 175 &   0.01  & 8.96 \\
 LWP12966  &   47252.94939  & 15  &   0.02  & 8.91 \\
 SWP33241  &   47260.78745  & 185 &   0.60  & 8.89 \\
 LWP13009  &   47260.93182  & 12  &   0.61  & 9.06 \\
 LWP13010  &   47260.97009  & 30  &   0.61  & 9.07 \\
 LWP13011  &   47261.01688  & 8   &   0.61  & 9.04 \\
 LWP28929  &   49583.48992  & 10  &   0.26  & *    \\
\hline
HIGH DISPERSION & & & & \\
\hline
 LWP08847  &   46648.93805  & 281  &  0.64  & 9.1 \\
 LWP12964  &   47252.84456  & 85   &  0.01  & 8.9 \\
 LWP12967  &   47253.23681  & 560  &  0.04  & 8.9 \\
 LWP12968  &   47253.53749  & 165  &  0.06  & 8.9 \\
 LWP13006  &   47260.56819  & 430  &  0.58  & 9.0 \\
\hline \hline
\multicolumn{5}{c}{HST Observations} \\
\hline
Observation ID & HJD & Grating & Texp  & Phase \\
               & -2400000 &    & (sec) &      \\
\hline
ob6b21020      &  55430.26348 & E230M & 1015 & 0.84 \\
ob6b21030      &  55430.27520 & E230M & 942 &  0.84 \\
ob6b21040      &  55430.34160 & E140M & 2917 & 0.84 \\
\hline

\end{tabular}
\end{center}
\end{table*}

\newpage
\begin{table*}
\caption{Parameters of the fit for each EPIC observation. \label{freepar}}
\begin{scriptsize}
\begin{tabular}{l c c c c c c c c}
\tableline\tableline
\noalign{\smallskip}
Obs. Id. & Phase & $N_\mathrm{H}$ & $Z/Z_\odot$ & k$T$  & $EM$ & $\chi^2_\mathrm{red}$ (d.o.f.) & Observed $f_\mathrm{X}$ & $\log L_\mathrm{X}$ \\
\noalign{\smallskip}
 & & ($\times 10^{21}$ cm$^{-2}$)  &  & (keV) & ($\times 10^{52}$ cm$^{-3}$)  & & ($\times 10^{-14}$ erg\,cm$^{-2}$\,s$^{-1}$) & (erg\,s$^{-1}$)\\
\noalign{\smallskip}
\tableline
\noalign{\smallskip}
\multicolumn{6}{l}{All parameters varying freely} \\
\noalign{\smallskip}
\tableline
\noalign{\smallskip}
0651870201 & 0.99 & 0.49$^{+0.64}_{-0.49}$ & 0.27$^{+0.16}_{-0.09}$ & 0.52$^{+0.06}_{-0.08}$ & 1.1$^{+1.1}_{-0.6}$ & 1.0 (58) & 6.5$^{+1.1}_{-6.2}$ & 28.9 \\
\noalign{\smallskip}
0651870301 & 0.15 & 0.08$^{+0.35}_{-0.08}$ & 0.26$^{+0.13}_{-0.09}$ & 0.58$^{+0.03}_{-0.04}$ & 1.0$^{+1.6}_{-0.2}$ & 1.0 (62) & 7.3$^{+0.9}_{-2.7}$ & 29.0 \\
\noalign{\smallskip}
0651870401 & 0.48 & 1.10$^{+0.68}_{-0.50}$ & 0.13$^{+0.05}_{-0.05}$ & 0.56$^{+0.05}_{-0.09}$ & 1.6$^{+1.2}_{-0.6}$ & 1.1 (44) & 4.4$^{+0.7}_{-4.0}$ & 28.7 \\
\noalign{\smallskip}
\tableline
\noalign{\smallskip}
\multicolumn{6}{l}{Global abundance and temperature fixed} \\
\noalign{\smallskip}
\tableline
\noalign{\smallskip}
0651870201 & 0.99 & 0.35$^{+0.29}_{-0.29}$ & $=$ 0.26 & $=$ 0.55 & 1.1$^{+0.2}_{-0.1}$ & 1.0 (60) & 6.6$^{+1.0}_{-0.9}$ & 28.9 \\
\noalign{\smallskip}
0651870301 & 0.15 & 0.23$^{+0.23}_{-0.26}$ & $=$ 0.26 & $=$ 0.55 & 1.1$^{+0.1}_{-0.1}$ & 1.0 (64) & 7.1$^{+0.9}_{-1.0}$ & 29.0 \\
\noalign{\smallskip}
0651870401 & 0.48 & 1.11$^{+0.49}_{-0.46}$ & $=$ 0.26 & $=$ 0.55 & 1.0$^{+0.2}_{-0.2}$ & 1.2 (46) & 4.2$^{+0.7}_{-0.9}$ & 28.7 \\
\noalign{\smallskip}
\tableline
\end{tabular}
\end{scriptsize}
\label{tab:freepar}
\end{table*}

\newpage
\begin{table*}
\caption{UV lines fluxes from IUE spectra}
\begin{scriptsize}
\begin{tabular}{l l l l l l l l }
\hline
Phase &   O~I &     C~II &  Si~IV & C~IV & Si~III & C~III & Mg~II \\
\tableline
      &\multicolumn{7}{c}{10$^{-13}$erg s$^{-1}$ cm$^{-2}$ }  \\
\hline
0.97&	$(2.8\pm 0.5)$& $(1.8 \pm 0.3)$ & $(2.3 \pm 0.5)$& $(5.5\pm 0.5)$
& $(3.5\pm 0.3)$ & $(2.1 \pm 0.2)$ & $(19 \pm 3)$  \\
$\sim 0.02$ & $(2.2\pm 0.4)$ & $(2.2 \pm 0.4)$ & $(2.3 \pm 0.6)$& $(5.9\pm 0.7)$
& $(2.7\pm 0.3)$ & $(1.8 \pm 0.2)$ & $(21 \pm 3)$ \\
$\sim 0.60$ & $(4.1\pm 0.9)$ & $(2.8 \pm 0.7)$ & $(2.6 \pm 1.2)$&$(5.6\pm 0.8)$
& $(3.1\pm 0.3)$ & $(2.4 \pm 0.3)$ & $(26 \pm 3)$ \\
\hline
\end{tabular}
\end{scriptsize}
\label{tab:uvflux}
\end{table*}

\newpage
\begin{table*}
\caption{Characteristics of the UV lines observed with HST/STIS}
\begin{tabular}{l l l l }
\hline
Line	& Wavelength	& Flux$^{(a)}$	& FWHM \\
	& \AA & $10^{-14}$erg~s~cm$^{-2}$ & km~s$^{-1}$\\
\hline
Si~III          &1206.5& 11.4$\pm 1.9$ & 400.2 \\
N~V 		&1238.8& 2.1$\pm 0.5$ & 278.7 \\
N~V		&1242.8& 0.6$\pm 0.3$  & - \\
O~I  	        &1302,1305,1306& 13.7$\pm 1.8$ & - \\
C~II		&1335,1336&13.8$\pm 0.9$  & - \\
Si~IV		&1393.8&8.2$\pm 1.1$  & 404.6 \\
Si~IV		&1402.8&9.5$\pm 1.2$ & 418.2\\
C~IV    	&1548,1550&35.4$\pm 2.7 $  & - \\
O~III		&1666&3.3		& - \\
Si~III		&1892.03&16.0	     & 219 \\
Mg~II		&2796,2803&$262. \pm 18 $ & - \\

\hline \hline
\multicolumn{4}{c}{H$_2$ lines} \\
\hline

P(2) 0-4 & 1338.554 & 0.35 & 56.1 \\ 
P(3) 0-4 & 1342.301 & 0.18 & 29.1 \\ 
P(2) 0-5 & 1399.013 & 0.29 & 34.0 \\ 
R(3) 1-6 (?) & 1430.918 & 0.26 & - \\ 
P(5) 1-7 & 1504.756 & 0.69 & 43.9 \\ 
P(8) 1-7 & 1524.597 & 0.27 & 31.5 \\ 
R(11) 2-8& 1556.803 & 0.19 & 28.2 \\ 
P(5) 1-8 & 1562.365 & 0.33 & 36.5 \\ 
\hline
& & & \\
& & & \\
$^{(a)}$ & \multicolumn{3}{l}{Fluxes are not extinction corrected.} \\

\end{tabular}
\end{table*}

\end{document}